    \documentclass[twocolumn,twocolappendix]{aastex631}
    \usepackage{newtxtext,newtxmath}
    \usepackage{amsmath}	% Advanced maths commands
    \usepackage{amssymb}	% Extra maths symbols
    \usepackage{footnote}
\hypersetup{linkcolor=red,citecolor=blue,filecolor=cyan,urlcolor=magenta}
\shorttitle{Realistic Cosmic Dust Model: and application to polarization of 2I/Borisov}
\shortauthors{Halder \& Sengupta}
\graphicspath{{./}{figures/}}
\begin{document}

%@arxiver{Figure_1.jpg,Figure_9.jpg}

\title{A comprehensive model of morphologically realistic Cosmic Dust particles: an application to mimic the unusual Polarization properties of the interstellar Comet 2I/Borisov}

\correspondingauthor{Prithish Halder}
\email{prithishh3@gmail.com, prithish.halder@iiap.res.in}

\author[0000-0002-1073-1419]{Prithish Halder}
\affiliation{Indian Institute of Astrophysics, Bangalore \\
Karnataka 560034, India}

\author[0000-0002-6176-3816]{Sujan Sengupta}
\affiliation{Indian Institute of Astrophysics, Bangalore \\
Karnataka 560034, India}

%\collaboration{6}{(AAS Journals Data Editors)}

%\author{Butler Burton}
%\affiliation{Leiden University}
%\affiliation{AAS Journals Associate Editor-in-Chief}

%\altaffiliation{AASTeX v6+ programmer}

%\affiliation{AAS Director of Publishing}
%\affiliation{American Astronomical Society \\
%1667 K Street NW, Suite 800 \\

%\author{Magaret Donnelly}
%\affiliation{IOP Publishing, Washington, DC 20005}



\begin{abstract}

The cosmic dust particles found in space are mainly porous aggregates of smaller grains. Theoretically, these aggregates are replicated using fractal geometry, assuming a cluster of spheres. Although, the light scattering response of cosmic dust aggregates has been thoroughly studied using clusters of spherical grains in the past few decades, yet, the effect of irregularities on the surface of each grain in an entire aggregate has mostly been neglected. We, for the first time, introduce a visually realistic cosmic dust model which incorporates a mixture of rough fractal aggregates (RFA) and agglomerated debris (Solids) to replicate the unusual polarization-phase curve observed in case of the interstellar comet 2I/Borisov at multiple wavelengths. The authenticity of the RFA structures has been verified by replicating light scattering results of circumstellar dust analogues from the Granada Amsterdam Light Scattering Database. We demonstrate that the light scattering response from the RFA structures has a very close resemblance with the experimental values. Finally, we model the observed polarization-phase curve of the interstellar comet 2I/Borisov using a mixture of RFA and solid particles. The best-fit data indicates presence of higher percentage of porous RFA structures (80\%) owing to the fact that the comet carries higher percentage of small and highly porous pristine cosmic dust particles. Further, the model indicates that the unusually steeper polarimetric slope and the high \textit{dust-to-gas} ratio in relatively newer comets is mainly due to higher \textit{porous-to-compact} ratio. 

\end{abstract}



\section{Introduction} \label{sec:intro}
The subject of interstellar comets is a very recent development in the field of astronomy, which started to emerge after the discovery of the Oort cloud. It was \cite{Sen1993} who for the first time predicted that one should detect an interstellar comet once in 200 years. The discovery of the interstellar comet 2I/Borisov in 2019, by Gennady Borisov after 180 years of cometary research has proved the above prediction to be true. Similar to other Solar System comets, 2I/Borisov exhibited a distinct coma allowing various researchers to study the physics and chemistry of the material content using spectroscopic and polarimetric observations. The spectroscopic studies of the comet 2I/Borisov indicate a \textit{dust-to-gas} ratio similar to those observed in carbon depleted comets of the Solar System \citep{Aravind2021ActivityObservatories, Yang2021Compact2I/Borisov}. On the other hand polarimetric observations indicate an unusually steeper slope \citep{Bagnulo2021}. Generally, Solar System comets are categorized in two polarimetric classes: low and high polarization comets depending on the different \textit{dust-to-gas} ratio observed in the coma \citep{Chernova1993, Levasseur-Regourd1996}. Apart from these two classes there exists a third class \citep{Hadamcik2003ImagingAngles} having polarization higher than that of high polarization comets, which was observed only in case one Solar System comet, C/1995 O1 (Hale-Bopp). Such high polarization is believed to be due to presence of extremely small Rayleigh size dust particles. The presence of Rayleigh size dust particles, in Hale-Bopp can be traced back to its origin in the outer regions of our Solar System, where the physical environment is comparable to that of interstellar medium. Similarly, 2I/Borisov is believed to have originated from the outer regions of its host stellar system. Although, Hale-Bopp might have visited the Sun once before its last apparition, 2I/Borisov on the hand, has not encountered any star before passing close to Sun, thereby the comet may hold a huge population of pristine cosmic dust particles. The interpretation and analysis of astronomical observations of dust in comets is mainly based on our knowledge of light scattering by morphologically irregular particles. The significance of dust particles having size comparable to the wavelength of incident light has been widely acknowledged \citep{Ahearn1995The1976-1992, AHearn2011EPOXI2, Kimura2006LightSpheres, Das2008AggregateHale-Bopp, Kolokolova2015PolarizationModel, Zubko2006DDAStructure, Zubko2020OnComets, DebRoy2017, Halder2018, Halder2021b}.  The cosmic dust particles found in space are mainly porous fractal aggregates of smaller grains formed due to coagulation and ballistic agglomeration in the circumstellar or interstellar environment. Theoretically, these aggregates are replicated using fractal geometry, assuming a cluster of spherical grains. But the studies of modelling of the third class of comets (Hale-Bopp) done by \cite{Lasue2006PorousObservations, Lasue2009Cometary1P/Halley} and \cite{Markkanen2015InhomogeneousDust} used aggregates of non-spherical monomers/grains and kept the monomer size fixed for multiple wavelengths. In order to verify whether 2I/Borisov holds relatively high amount of pristine cosmic dust particles or to have an estimate of the amount of pristine dust present in the coma of the comet, it is necessary to conduct light scattering simulations over modelled pristine cosmic dust aggregates and replicate the observed unusual polarization with exact computer modelled replica of cosmic dust. The dust particles studied by the Rosetta/MIDAS and Rosetta/COSIMA suggest presence of porous aggregated dust particles that resemble the morphology of interplanetary dust (IDP) \citep{Bentley2016Aggregate67P/ChuryumovGerasimenko, Guttler2019Synthesis67P/Churyumov-Gerasimenko, Mannel2019, Schulz2015CometYears}. The IDPs collected from the Earth's stratosphere and Antarctic ice having irregular geometry, fluffy aggregates and fractal nature represents the physical morphology of Solar system cosmic dust \citep{Brownlee2003COSMICRESEARCH, Noguchi2015CometarySurface}. Again, due to flash heating in the upper atmosphere, these IDP samples may not purely represent pristine cosmic dust. The cosmic dust analog aggregates prepared in the Granada Amsterdam Light Scattering facility using Condensation Flow Apparatus shall represent the most pristine morphology of cosmic dust which are devoid of flash heating. The microgravity and laboratory experiments of dust-dust interactions conducted to replicate the conditions prevailing in the early Solar System suggest formation of fractal assemblage of dust via ballistic agglomeration \citep{Blum2000, Krause2004, Wurm1998}. In a similar way, small dust particles in the interstellar medium may coagulate in the vicinity of dense molecular clouds. Hence, astronomers around the world use fractal aggregates/cluster of spheres to study the physical and/or optical properties of cosmic dust. Numerically fractal aggregates are prepared using ballistic agglomeration techniques. These agglomeration techniques hold the physics behind the dust coagulation in circumstellar and protoplanetary disks, but the morphology of each grain in an aggregate lacks surface roughness or irregularities. Roughness has been a matter of concern for a longer period of time and hence, dust structures such as gaussian random sphere, agglomerated debris and rough spheroids are developed by various researchers \citep{Kolokolova2015PolarizationModel, Muinonen1996LightApproximation, Zubko2006DDAStructure} to include the contribution of irregularities or roughness. Although, these rough or irregular structures explain the contribution of roughness in case of single particles and debris particles, the contribution of roughness or irregularities on the surface of each grain of a fractal aggregate remains unknown. 

In the present investigation, for the first time, we use a visually realistic cosmic dust model which is represented by a mixture of highly porous Rough Fractal Aggregate (RFA) \cite{Halder2022a} and low porous Solids (Agglomerated Debris) to model the unusual polarization properties of the interstellar comet 2I/Borisov. The highly porous RFA structures which are aggregates of irregular/rough grains have a very close resemblance with the IDPs collected from Earth's stratosphere. Initially, the RFA modelled dust particles are validated by replicating light scattering results from Granada Amsterdam Light Scattering Database for the different aggregates samples (1-6)\citep{Volten2007AstrophysicsAnalogs} of circumstellar or cosmic dust analogs \citep{Nuth2000, Rietmeijer1999}. Then, we model the observed polarization-phase curve and the polarimetric spectral gradient of the interstellar comet 2I/Borisov using a mixture of RFA model structures (high porous) and Solids (low porous) at the three wavelengths, $\lambda$ = 0.557$\mu$m ($V_f$ filter), 0.655$\mu$m ($R_f$ filter) and 0.768$\mu$m ($I_f$ filter) respectively. Finally, we compare the observed \textit{dust-to-gas} ratio with the intrinsic dust parameter \textit{porous-to-compact} ratio for the extremely high-polarization comets (2I/Borisov \& Hale-Bopp) and low-polarization comets (67P/C-G \& 1P/Halley) to understand the dependence of dust-to-gas ratio on the intrinsic dust parameters.

\section{Modelling Methodology} \label{sec:model_methods}
In this section, we describe the techniques employed to generate RFA (Rought Fractal Aggregates) and Solid particles used to replicate the light scattering results from the Granada Amsterdam Light Scattering Database and to model the polarization properties of the of the interstellar comet 2I/Borisov. The light scattering technique and the related light scattering parameters are also discussed in this section.

\subsection{Fractal Aggregates (FA)} \label{sec:fractal_agg}
\begin{figure*}[ht]
    \centering
   % \vspace{10cm}
    \includegraphics[scale=0.115]{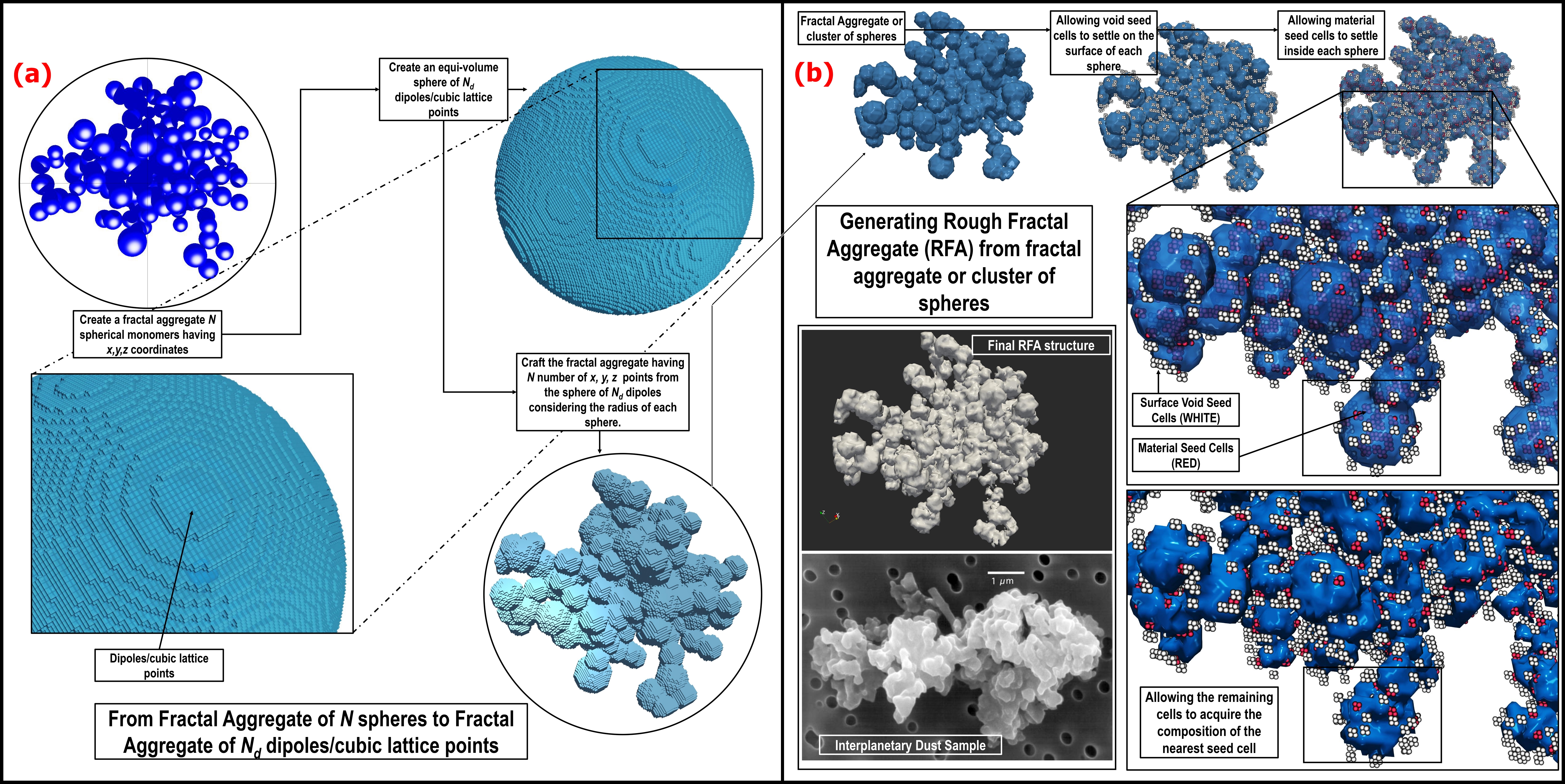}
    \caption{(a)Algorithm to generate Fractal Aggregates (FA) having \textit{N$_d$} dipoles and (b) algorithm to generate Rough Fractal Aggregates (RFA) using REST \citep{Halder2022a}.}
    \label{fig:1}
\end{figure*}

Fractal aggregates (FA) having polydisperse spheres are created following the BPCA and BCCA agglomeration techniques using the Java package FLAGE\footnote{FLAGE \url{https://scattering.eu/}} \citep{Skorupski2014}. The structure of the FA is loaded in the package REST \citep{Halder2022a} where the x, y, z coordinates and radii of each sphere of an aggregate are scaled into an equi-volume sphere made up of unit dipoles/lattice points. In REST the RFA algorithm remove those dipoles/lattice points which do not fall within the radii of each sphere and hence forming the resultant structure which is a fractal aggregate (FA) of spheres but made up of dipoles/lattice points as shown in Figure-\ref{fig:1}(a). REST is a structure tool that generates realistic cosmic dust particles from spheres, superellipsoids and fractal aggregates (FA). 
FLAGE is a very useful java tool to create aggregates of spherical grains. It takes the following physical parameters as input to create a proper fractal aggregate:

\begin{enumerate}
    \item Number of spheres (\textit{N}).
    \item Radius of each sphere/primary particle (r$_{p}$).
    \item Radius of aggregate (\textit{R$_{a}$ = $\sqrt{5/3}$R$_{g}$, where R$_{a}$ is the characteristic radius of the aggregate and R$_g$ is the radius of gyration which is defined in equation 1}).
    \item Fractal dimension (\textit{D$_{f}$, a dimensionality constant that characterises a fractal structure and is defined by equation 1}).
    \item Fractal prefactor (\textit{k$_{f}$, a proportionality constant that is a prefactor of the fractal scaling relation defined by equation 1}).
    \item Porosity (\textit{The degree or percentage of space present within a fractal aggregate}).
    \item Aggregate Type:
    \begin{enumerate}
        \item Ballistic Particle Cluster Agglomeration (\textit{BPCA}).
        \item Ballistic Cluster-Cluster Agglomeration (\textit{BCCA}).
        \item Diffusion Limited Agglomeration (\textit{DLA}).
        \item Reaction Limited Agglomeration (\textit{RLA}).
    \end{enumerate}
    
\end{enumerate}

The structural arrangement of an aggregate having $N$ monodisperse (spheres having same size) spherical grains (each having radius $r$) is defined by the following equation \citep{Sorensen1992Light-scatteringFlames},

\begin{equation}
    N = k_{f}\left(\frac{R_{g}}{r}\right)^{D_{f}}
\end{equation}

But the aggregates found in space are polydisperse (spheres having different sizes) in nature. An aggregate having polydisperse spherical grains with primary particle (PP) radius $r_{p}$, average PP mass $\overline{m_p}$ and aggregate mass $m_a$ is defined by \cite{Eggersdorfer2011TheParticles} as

\begin{equation}
    \frac{m_a}{\overline{m_p}} = k_{f}\left(\frac{R_{g}}{r_p}\right)^{D_{f}}
\end{equation}

\noindent where $m_a/\overline{m_p} = N$ is the number of PP in the aggregate.\\

The fractal dimension is directly related to the porosity of an aggregate. The BCCA aggregates have porosity $\geq$ 95\% and \textit{D$_f$} $<$ 2, while the BPCA aggregates have porosity $\leq$ 90\% and \textit{D$_f$} $>$ 2. The porosity of different FA structures is controlled by changing \textit{D$_f$} = 1.8 to 2.5. This is done to incorporate minute variation in porosity following the explanations provided for the different aggregate samples from the Granada Amsterdam Light Scattering Database. The aggregate samples 1 and 2 \citep{Volten2007AstrophysicsAnalogs} are made up of the same material, with the same grain size and the same aggregate size, yet the polarization maximum (\textit{P$_{max}$}) for the two samples are different. It is already clear from previous studies that a slight increase in porosity shall induce an increase in the \textit{P$_{max}$} value \citep{Kimura2006LightSpheres, Halder2018}. Also, the authors of the Granada Amsterdam Light Scattering Database have mentioned that this shift in \textit{P$_{max}$} for the aggregate samples 1 and 2 is possibly due to minute difference in porosity.

\subsection{Rough Fractal Aggregates (RFA)}
The Rough Fractal Aggregate (RFA) structures used in this study are generated using the java package REST\footnote{Rough Ellipsoid Structure Tools (REST) \url{https://rest-package.readthedocs.io/}} (Rough Ellipsoid Structure Tools) \citep{Halder2022a} from the loaded FA structure file. The RFA algorithm in REST crafts surface roughness/irregularities on the surface of each spherical grain of a FA structure [see Figure-\ref{fig:1}(b)]. The algorithm to generate the RFA structures is discussed below:

\begin{enumerate}
    \item Browse and select the structure file of a fractal aggregate/cluster of spheres having following format ($i$, $R$, $X$ $Y$ $Z$, \emph{mtag}, \emph{mtag}), where $i$ is the sphere number, $R$ is the $i^{th}$ sphere radius (in $\mu$m), $X$, $Y$, $Z$ are the coordinates of each sphere (monomer) and the \emph{mtag} is the composition tag.
    
    \item Multiply each coordinate and radii with an integer scale factor $n$. This done to achieve a desired number of dipoles (\textit{N$_d$}) for the entire RFA structure.
    
    \item Measure the distance ($d_{mono}$) of each monomer from the centre ($0$,$0$,$0$).

    \item Create the initial sphere having $N_{d}$ dipoles and radius $R_d$ = \texttt{maximum}($d_{mono}$) (in number of dipoles) from the centre of the initial sphere.
    
    \item Randomly choose $N_{ss}$ surface seed cells on the surface of the FA inside the initial sphere. 
    
    \item Randomly choose $N_{m}$ material seed cells inside the surface of the FA.
    
    \item Measure the distance $D_{m}^{i,j}$ between the $j$th material seed cell and $i$th dipole of the base structure, where $j$ = 1 to $N_{m}$ and $i$ = 1 to $N_{d}$.
    
    \item Measure the distance $D_{ss}^{i,k}$ between the $k$th surface space seed cell and $i$th dipole of the base structure, where $k$ = 1 to $N_{ss}$ and $i$ = 1 to $N_{d}$.
    
    \item Print those dipoles for which, $D_{m}^{i}$ $<$ $D_{ss}^{i}$ in the final RFA structure file.

\end{enumerate}

The porosity $\mathrm{P}$ of a fractal aggregate is determined by the ratio of total number of space seed cells in entire volume of sphere circumscribing the RFA structure by the total volume of the circumscribing sphere \citep{Halder2022a},
\[\texttt{Volume of initial structure} (V_i) = N_d(\texttt{initial})\times d^3\]
\[\texttt{Volume of final structure} (V_f)   = N_d(\texttt{final})\times d^3\]

The total volume of space seed cells is given by,
\[V_T = [N_d(\texttt{final}) - N_d(\texttt{initial})]\times d^{3}\] 

Therefore, the degree of porosity is,
\begin{equation}
\begin{split}
    \mathrm{P} & = \frac{V_T}{V_f} = \frac{[N_d(\texttt{final}) - N_d(\texttt{initial})]\times d^{3}}{N_{d}(\texttt{final})\times d^{3}} \\
    & = \frac{[N_d(\texttt{final}) - N_d(\texttt{initial})]}{N_{d}(\texttt{final})}
\end{split}
\end{equation}

\subsection{Generating Solid structures}
The Solid structures used in this study are low porous Agglomerated Debris particles generated using REST following the Poked Structure (PS) option/algorithm. The steps to generate PS shape are as follows:

\begin{enumerate}
%    \item Follow steps 1 to 2 from the SDS algorithm in Section-\ref{subsubsection:sds}.
    \item Generate initial spherical structure file \texttt{target.out} having $N_{d}$ dipoles and radius $R$ (in number of dipoles) using \texttt{CALLTARGET} module.

    \item Randomly choose $N_{m}$ number of material seed cells from the $N_{d}$ dipoles present in the \texttt{target.out} file.
    
    \item Randomly choose $N_{is}$ number of internal space seed cells from the $N_{d}$ dipoles present in the \texttt{target.out} file.
        
    \item Randomly choose $N_{ss}$ number of surface space seed cells from the $N_{d}$ dipoles present in the \texttt{target.out} file. The surface thickness should be $t$ times $r$.
    
    \item Measure the distance $D_{m}^{i,j}$ between the $j$th material seed cell and $i$th dipole of the base structure, where $j$ = 1 to $N_{m}$ and $i$ = 1 to $N_{d}$.
    \item Measure the distance $D_{is}^{i,k}$ between the $k$th internal space seed cell and $i$th dipole of the base structure, where $k$ = 1 to $N_{is}$ and $i$ = 1 to $N_{d}$.
    \item Measure the distance $D_{ss}^{i,l}$ between the $l$th surface space seed cell and $i$th dipole of the base structure, where $l$ = 1 to $N_{ss}$ and $i$ = 1 to $N_{d}$.
   \item Print those dipoles for which, $D_{m}^{i}$ $<$ $D_{is}^{i}$ and $D_{m}^{i}$ $<$ $D_{ss}^{i}$ in the final structure file.

\end{enumerate}

The steps to generate Solid structures (AD particles) using PS algorithm are:
\begin{itemize}
    
    \item $R_d$ (in number of dipoles) = 64
    \item $N_m$ = 21
    \item $N_{is}$ = 20
    \item $N_{ss}$ = 100
    \item $t$ = 1\%

\end{itemize}

\subsection{Light scattering simulations}
The coma of a comet is optically thin having low volume concentration of dust particles. Hence, in theoretical modelling of the light scattering by the dust particles in the coma of comet, multiple scattering effects are neglected. The scattering  phenomenon for a mirror symmetric and macroscopically isotropic particulate medium is defined by the scattering matrix which represents far field transformation of the Stokes parameters of the incident light ($I_i$, $Q_i$, $U_i$, $V_i$) to that of the scattered light ($I_s$, $Q_s$, $U_s$, $V_s$).   This scattering matrix is given by \cite{Bohren1998AbsorptionParticles}:
\begin{equation}
\left( \begin{array}{cccc}
I_s  \\
Q_s  \\
U_s  \\
V_s  \end{array} \right)= \frac{1}{k^{2}d^{2}} \left( \begin{array}{cccc}
    S_{11} & S_{12} & 0 & 0 \\
    S_{12} & S_{22} & 0 & 0 \\
    0 & 0 & S_{33} & S_{34} \\
    0 & 0 & -S_{34} & S_{44} \end{array} \right)\left( \begin{array}{cccc}
I_i  \\
Q_i  \\
U_i  \\
V_i  \end{array} \right)
\end{equation}
where $k$ is the wave-number and $d$ is the distance between the scatterer and the observer and $S_{ij}$ represents the orientationally symmetric scattering matrix elements. The angle $\alpha$ between Sun-Comet-Earth is called the \emph{Phase angle}. Angle $\theta$ = 180 - $\alpha$ is called the \emph{Scattering angle}, $\theta$ = [0$^{\circ}$,180$^{\circ}$]. 

In this work, we study the following light scattering parameters defined by the scattering matrix elements:

\begin{enumerate}
    \item \emph{Phase function}: $S_{11}$  
    
    \item \emph{Degree of linear polarization}: \(DP = -S_{12}/S_{11}\).
    
    \item \(\frac{S_{22}}{S_{11}}\)
    
\end{enumerate}
   
The anisotropy condition for a non spherical scatterer is, \(S_{11} \neq S_{22}\) and \(S_{33} \neq S_{44}\).

We use the Discrete Dipole Approximation(DDA) scattering codes\footnote{Discrete Dipole Approximation (DDA) DDSCAT version 7.3.3 \url{http://ddscat.wikidot.com/}} {\cite{draine1994}} in parallel mode, for the numerical simulations of light scattering.

Further, the results are averaged using the power-law size distribution $r^{-n}$ where $n$ ranges between 2.0 to 3.0. The power-law size distribution is modelled by considering aggregates of different sizes from smallest to largest. The aggregate sizes are increased by increasing the number of monomers/grains and keeping the monomer size fixed.  

\begin{table*}
%\begin{center}
%\begin{minipage}{\textwidth}
\caption{Model$_X$ RFA parameters: details of different parameters of RFA structures such as number of monomers (\textit{N}), aggregate size parameter (\textit{X$_a$}), fractal dimension (\textit{D$_f$}), monomer radius (\textit{r$_p$}) and aggregate radius (\textit{R$_a$}) in the three different wavelengths for the minimum and maximum sizes respectively.}\label{tab1}
\scriptsize
\begin{tabular*}{\textwidth}{@{\extracolsep{\fill}}lccccccccc@{\extracolsep{\fill}}}
\hline
%& \multicolumn{6}{@{}c@{}}{Aggregate Samples 1\footnotemark[1]} & \multicolumn{6}{@{}c@{}}{RFA Model structures 2\footnotemark[2]} \\\cmidrule{2-7}\cmidrule{8-13}%
\textit{N} & \textit{X$_a$} & \textit{D$_f$} & \textit{r$_p$}($\lambda=0.557\micron$) & \textit{R$_a$}($\lambda=0.557\micron$) & \textit{r$_p$}($\lambda=0.655\micron$) & \textit{R$_a$}($\lambda=0.655\micron$) & \textit{r$_p$}($\lambda=0.768\micron$) & \textit{R$_a$}($\lambda=0.768\micron$)  \\
\hline
\hline
45	&	5	& 1.8 &	0.062$\mu$m	&	0.44$\mu$m	&	0.073$\mu$m	&	0.52$\mu$m & 0.085$\mu$m & 0.61$\mu$m\\
625	&	21  & 1.8 &	0.062$\mu$m &	1.91$\mu$m	&	0.073$\mu$m	&	2.25$\mu$m & 0.085$\mu$m & 2.62$\mu$m \\
\hline

\end{tabular*}
%\footnotetext{Note: This is an example of table footnote. This is an example of table footnote this is an example of table footnote this is an example of~table footnote this is an example of table footnote.}
%\footnotetext[1]{The different samples of circumstellar dust analog prepared in the Granada Amsterdam Light Scattering setup is described in Table-3 of \cite{Volten2007AstrophysicsAnalogs}.}
%\footnotetext[2]{The different realizations of the RFA Model structures are shown in Figure-\ref{fig:A1} in Appendix-A.}
%\end{minipage}
%\end{center}
\end{table*}

\begin{table*}
%\begin{center}
%\begin{minipage}{\textwidth}
\caption{Model$_X$ Solid parameters: details of different parameters of Solid structures such as size parameter (\textit{X}), packing fraction (\textit{$\rho_f$}) and radius (\textit{R}) in the three different wavelengths for the minimum and maximum sizes respectively.}\label{tab2}
\scriptsize
\begin{tabular*}{\textwidth}{@{\extracolsep{\fill}}lcccccccc@{\extracolsep{\fill}}}
\hline
%& \multicolumn{6}{@{}c@{}}{Aggregate Samples 1\footnotemark[1]} & \multicolumn{6}{@{}c@{}}{RFA Model structures 2\footnotemark[2]} \\\cmidrule{2-7}\cmidrule{8-13}%
\textit{X} & $\rho_f$ & \textit{R}($\lambda=0.557\micron$) & \textit{R}($\lambda=0.655\micron$) & \textit{R}($\lambda=0.768\micron$)\\
\hline
\hline
0.65 & 0.26	& 0.057$\mu$m	& 0.067$\mu$m &	0.079$\mu$m\\
20	 & 0.27 & 1.77$\mu$m   & 2.08$\mu$m & 2.44$\mu$m\\
\hline

\end{tabular*}
%\footnotetext{Note: This is an example of table footnote. This is an example of table footnote this is an example of table footnote this is an example of~table footnote this is an example of table footnote.}
%\footnotetext[1]{The different samples of circumstellar dust analog prepared in the Granada Amsterdam Light Scattering setup is described in Table-3 of \cite{Volten2007AstrophysicsAnalogs}.}
%\footnotetext[2]{The different realizations of the RFA Model structures are shown in Figure-\ref{fig:A1} in Appendix-A.}
%\end{minipage}
%\end{center}
\end{table*}

\begin{table*}
%\begin{center}
%\begin{minipage}{\textwidth}
\caption{Model$_R$ RFA parameters: details of different parameters of RFA structures such as number of monomers (\textit{N}), fractal dimension (\textit{D$_f$}), monomer radius (\textit{r$_p$}), aggregate radius (\textit{R$_a$}) and aggregate size parameter (\textit{X$_a$}) in the three different wavelengths for the minimum and maximum sizes respectively.}\label{tab3}
\scriptsize
\begin{tabular*}{\textwidth}{@{\extracolsep{\fill}}lcccccccc@{\extracolsep{\fill}}}
\hline
%& \multicolumn{6}{@{}c@{}}{Aggregate Samples 1\footnotemark[1]} & \multicolumn{6}{@{}c@{}}{RFA Model structures 2\footnotemark[2]} \\\cmidrule{2-7}\cmidrule{8-13}%
\textit{N} & \textit{D$_f$} & \textit{r$_p$} & \textit{R$_a$} & \textit{N$_d$} & \textit{X$_a$}($\lambda=0.557\micron$) & \textit{X$_a$}($\lambda=0.655\micron$) & \textit{X$_a$}($\lambda=0.768\micron$)\\
\hline
\hline
10	&	1.8 & 0.073$\mu$m	& 0.22 & 1,675 &	2.48	&	2.11	&	1.79   \\
625	&	1.8 & 0.073$\mu$m & 2.00 & 90,780 & 	22.5	&	19.18	&	16.36  \\
\hline

\end{tabular*}
%\footnotetext{Note: This is an example of table footnote. This is an example of table footnote this is an example of table footnote this is an example of~table footnote this is an example of table footnote.}
%\footnotetext[1]{The different samples of circumstellar dust analog prepared in the Granada Amsterdam Light Scattering setup is described in Table-3 of \cite{Volten2007AstrophysicsAnalogs}.}
%\footnotetext[2]{The different realizations of the RFA Model structures are shown in Figure-\ref{fig:A1} in Appendix-A.}
%\end{minipage}
%\end{center}
\end{table*}

\begin{table*}
%\begin{center}
%\begin{minipage}{\textwidth}
\caption{Model$_R$ Solid parameters: details of different parameters of Solid structures such as radius (\textit{R}), packing fraction (\textit{$\rho_f$}) and size parameter (\textit{X}) in the three different wavelengths for the minimum and maximum sizes respectively.}\label{tab4}
\scriptsize
\begin{tabular*}{\textwidth}{@{\extracolsep{\fill}}lcccccccc@{\extracolsep{\fill}}}
\hline
%& \multicolumn{6}{@{}c@{}}{Aggregate Samples 1\footnotemark[1]} & \multicolumn{6}{@{}c@{}}{RFA Model structures 2\footnotemark[2]} \\\cmidrule{2-7}\cmidrule{8-13}%
\textit{R$_a$} & \textit{$\rho_f$} & \textit{X}($\lambda=0.557\micron$) & \textit{X}($\lambda=0.655\micron$) & \textit{X}($\lambda=0.768\micron$)\\
\hline
\hline
0.07$\mu$m & 0.26 & 0.79	& 0.67 &	0.57\\
2.0$\mu$m  & 0.27 & 22.5   & 19.2 & 16.4\\
\hline

\end{tabular*}
%\footnotetext{Note: This is an example of table footnote. This is an example of table footnote this is an example of table footnote this is an example of~table footnote this is an example of table footnote.}
%\footnotetext[1]{The different samples of circumstellar dust analog prepared in the Granada Amsterdam Light Scattering setup is described in Table-3 of \cite{Volten2007AstrophysicsAnalogs}.}
%\footnotetext[2]{The different realizations of the RFA Model structures are shown in Figure-\ref{fig:A1} in Appendix-A.}
%\end{minipage}
%\end{center}
\end{table*}

\begin{figure}
    \centering
   % \vspace{10cm}
    \includegraphics[scale=0.35]{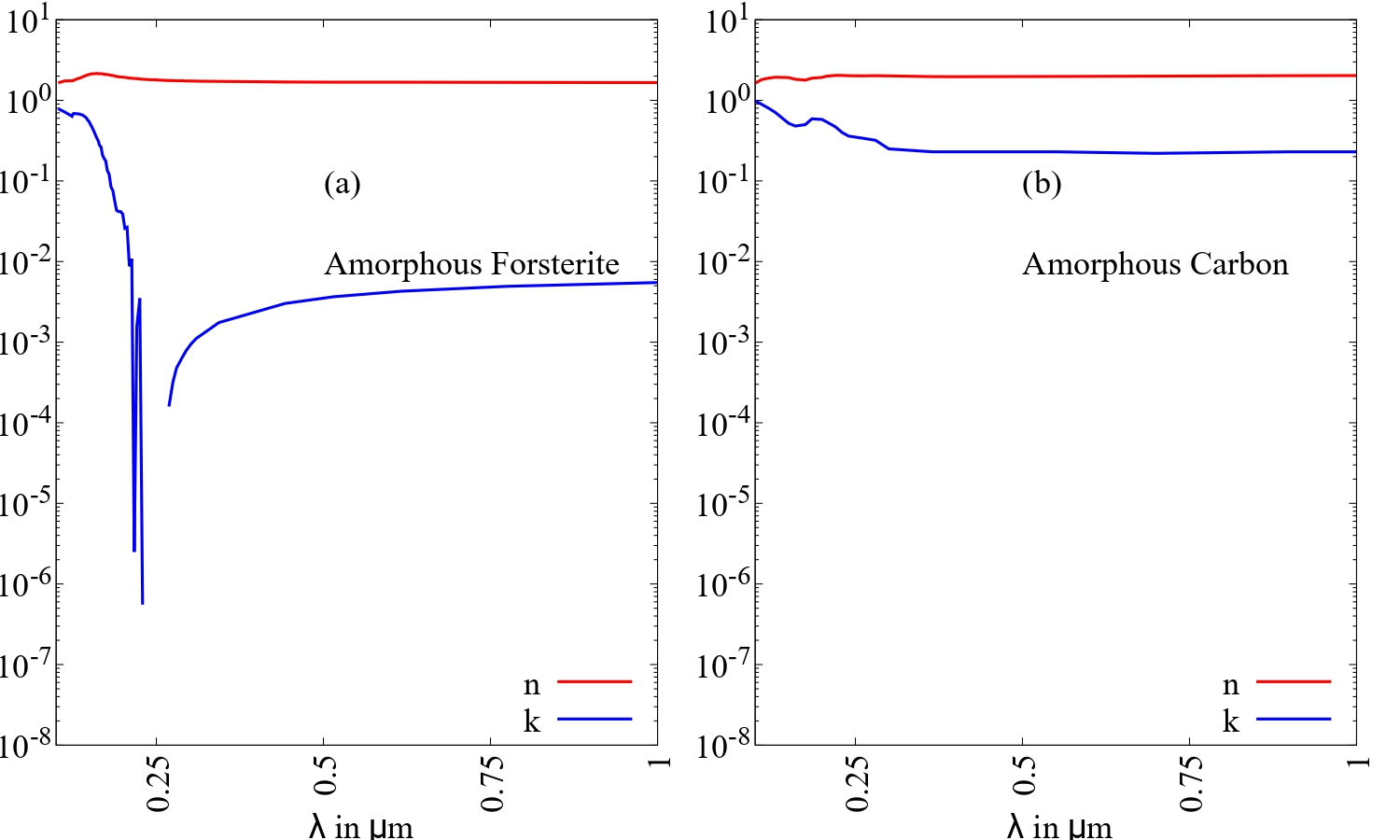}
    \caption{Variation of complex refractive index \textit{n} (red line) and \textit{k} (blue line) with increasing wavelength for (a) amorphous forsterite and (b) amorphous carbon.}
    \label{fig:2}
\end{figure}

\subsection{Dust Model} \label{sec:dustmodel}
In this study, we have considered two kinds of modelling approaches to extract the best possible results. The first modelling approach considers fixed values of monomer size parameters (x) for all the three wavelengths and we term it as Model$_{X}$. While the second modelling approach considers fixed values of monomer size (r) for all the three wavelengths and we term it as Model$_{R}$. In both models we use a mixture of highly porous RFA structures and Solid agglomerated debris particles. 
As silicates and carbonaceous materials majorly constitute the composition of dust found in comets \citep{Bardyn2017Carbon-richCOSIMA/Rosetta}, we have considered the refractive indices of amorphous forsterite to represent silicates and amorphous carbon to represent carbonaceous materials. We have considered the refractive indices of amorphous forsterite to represent silicates and amorphous carbon to represent carbonaceous materials. Figure-\ref{fig:2} shows the wavelength dependence of refractive index for both amorphous forsterite \citep{Scott1996UltravioletSilicates} and amorphous carbon \citep{Jenniskens1993OpticalResidue, Li1997ADust.}, while the specific values of refractive indices in the three wavelengths for silicate and carbon are shown below. 

  \[
    Silicate\left\{\begin{array}{lr}
        1.68 + 0.0035i, & \text{for } \lambda = 0.557~\micron\\
        1.67 + 0.0040i, & \text{for } \lambda = 0.655~\micron\\
        1.66 + 0.0048i, & \text{for } \lambda = 0.768~\micron
        \end{array}\right\}
  \]

  \[
    Carbon\left\{\begin{array}{lr}
        1.97 + 0.23i, & \text{for } \lambda = 0.557~\micron\\
        1.99 + 0.22i, & \text{for } \lambda = 0.655~\micron\\
        2.0 +  0.2i, & \text{for } \lambda = 0.768~\micron
        \end{array}\right\}
  \]  

The details of both the modelling approaches are explained in the following sub-sub-sections.

\begin{table*}
\begin{center}
\begin{minipage}{\textwidth}
\caption{Physical properties of RFA Model structures (1-6)}\label{tab5}
\scriptsize
\begin{tabular*}{\textwidth}{@{\extracolsep{\fill}}lccccccc@{\extracolsep{\fill}}}
\hline
%& \multicolumn{6}{@{}c@{}}{Aggregate Samples 1\footnotemark[1]} & \multicolumn{6}{@{}c@{}}{RFA Model structures 2\footnotemark[2]} \\\cmidrule{2-7}\cmidrule{8-13}%

Sample\footnote{The different samples of circumstellar dust analog prepared in the Granada Amsterdam Light Scattering setup is described in Table-3 of \cite{Volten2007AstrophysicsAnalogs}.} & RFA Structures\footnote{The different realizations of the RFA Model structures are shown in Figure-\ref{fig:A1} in Appendix-A.} & $n$\footnote{For simplicity we considered the imaginary part of the refractive index $k$ = 0.0001 for all the structures, as it was not detected in the experiments.} \& colour\footnote{The colour refers to the observed colours of different samples from the experiments.} & porosity & grain radius & aggregate radius (\textit{R$_a$}) & Average no. of dipoles (\textit{N$_d$})\\
\hline
\hline
1	&	1.1, 1.2, 1.3	&	1.7 dark brown	&	91\% - 95\%	&	0.05-0.12$\mu$m	&	0.65$\mu$m & 25,896\\
2	&	2.1, 2.2    	&	1.7 dark brown &	93\% - 94\%	&	0.05-0.12$\mu$m	&	0.65$\mu$m & 24,500\\
3	&	3.1, 3.2, 3.3	&	1.6	light brown   &	94\% - 95\%	&	0.03-0.05$\mu$m	&	0.65$\mu$m & 25,423\\
4	&	4.1, 4.2, 4.3	&	1.6	light brown   &	93\% - 94\%	&	0.03-0.05$\mu$m	&	0.6$\mu$m & 18,809\\
5	&	5.1, 5.2, 5.3	&	1.8 black        &	97\% - 99\%	&	0.015-0.06$\mu$m	&	0.375$\mu$m & 3,305\\
6	&	6.1, 6.2, 6.3	&	1.7 black        &	91\% - 96\%	&	0.015-0.06$\mu$m	&	0.6$\mu$m & 19,324\\
\hline

\end{tabular*}
%\footnotetext{Note: This is an example of table footnote. This is an example of table footnote this is an example of table footnote this is an example of~table footnote this is an example of table footnote.}
%\footnotetext[1]{ The different samples of circumstellar dust analog prepared in the Granada Amsterdam Light Scattering setup is described in Table-3 of \cite{Volten2007AstrophysicsAnalogs}.} 
%\footnotetext[2]{ The different realizations of the RFA Model structures are shown in Figure-\ref{fig:A1} in Appendix-A.}

%\footnotetext[3]{ For simplicity we considered the imaginary part of the refractive index $k$ = 0.0001 for all the structures, as it was not detected in the experiments.} 

%\footnotetext[4]{ The colour refers to the observed colours of different samples from the experiments.}

\end{minipage}
\end{center}
\end{table*}

\begin{figure*}
    \centering
    \vspace{-1cm}
    \includegraphics[scale=0.2]{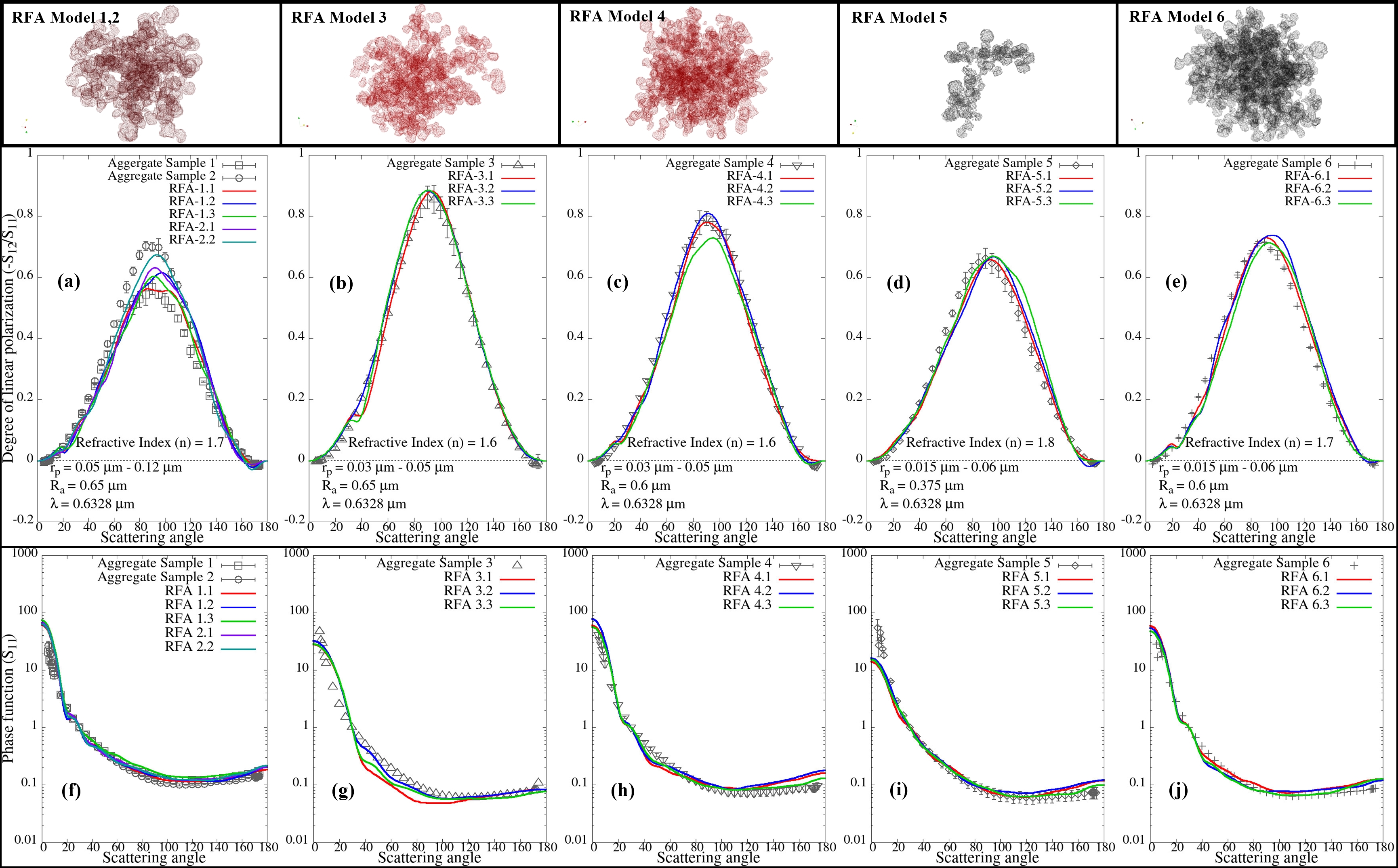}
    \caption{Variation of the degree of linear polarization \textit{-S$_{12}$/S$_{11}$} (a-e) and phase function \textit{S$_{11}$} (f-j) with scattering angle for RFA model structures (1-6) (Solid lines) compared with the experimental results for Aggregate Samples (1-6) (hollow squares and triangles) from the Granada Amsterdam Light Scattering Database \citep{Volten2007AstrophysicsAnalogs}.}
    \label{fig:3}
\end{figure*}

\begin{figure*}
    \centering
    %\vspace{-5cm}
    \includegraphics[scale=0.8]{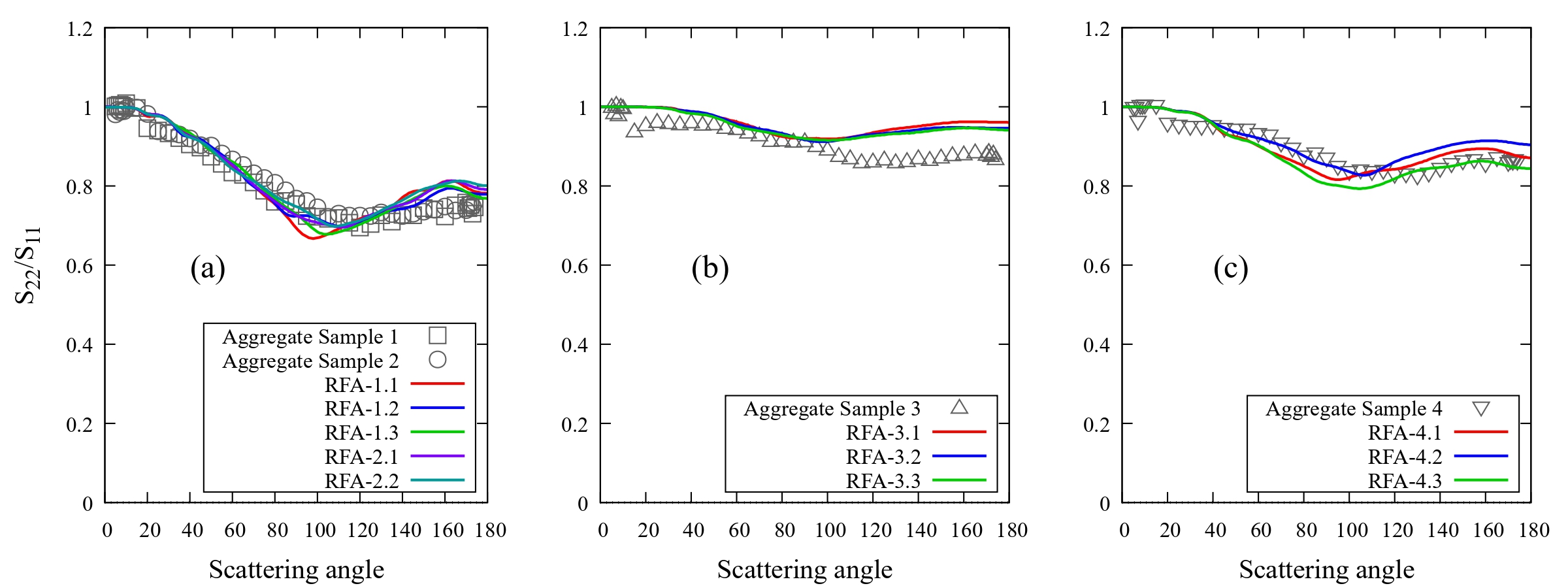}
    \caption{Variation of \textit{S$_{22}$/S$_{11}$} (a-c) with scattering angle for RFA model structures (1-4) (Solid lines) compared with the experimental results for Aggregate Samples (1-4) (hollow squares and triangles) from the Granada Amsterdam Light Scattering Database \citep{Volten2007AstrophysicsAnalogs}.}
    \label{fig:4}
\end{figure*}  

\subsubsection{Model$_X$: constant size parameter}
In this modelling approach we consider fixed primary particle (monomer) size parameter (\textit{x$_{p}$} = 0.7) for the RFA model structures which corresponds to following PP (monomer) radii ($r_p$),

  \[
    x_p = \frac{2\pi}{\lambda}r_p = 0.7\left\{\begin{array}{lr}
        r_p = 0.062\micron, & \text{for } \lambda = 0.557~\micron\\
        r_p = 0.073\micron, & \text{for } \lambda = 0.655~\micron\\
        r_p = 0.085\micron, & \text{for } \lambda = 0.768~\micron
        \end{array}\right\}
  \]
The upper and lower size cuttoffs of monomer size under Model$_X$ is defined using lognormal size distribution having standard deviation of $\sigma_p$ = $\pm$ 0.03$\micron$. Thus the smallest monomer sizes are 0.032$\micron$, 0.043$\micron$ and 0.055$\micron$ at $V_f$, $R_f$ and $I_f$ filters respectively. While the largest monomer sizes are 0.092$\micron$, 0.103$\micron$ and 0.115$\micron$ at $V_f$, $R_f$ and $I_f$ filters respectively.
In total, 50 RFA aggregates are considered having aggregate size parameters ($X_a$) in the range 5 to 20 and number of monomers ($N$) in the range 45 to 625 (see Table-1 for more details).

On the other hand, for the low porosity solid particles we have considered agglomerated debris particles generated ussing REST. A total of 50 solid particles are considered having minimum size parameter 0.65 and maximum size parameter 20. The respective minimum and maximum radii of the solids for the three wavelengths are depicted below,

  \[
    X_{min} = \frac{2\pi}{\lambda}R = 0.65\left\{\begin{array}{lr}
        R = 0.057\micron, & \text{for } \lambda = 0.557~\micron\\
        R = 0.067\micron, & \text{for } \lambda = 0.655~\micron\\
        R = 0.079\micron, & \text{for } \lambda = 0.768~\micron
        \end{array}\right\}
  \]

    \[
    X_{max} = \frac{2\pi}{\lambda}R = 20\left\{\begin{array}{lr}
        R = 1.77\micron, & \text{for } \lambda = 0.557~\micron\\
        R = 2.08\micron, & \text{for } \lambda = 0.655~\micron\\
        R = 2.44\micron, & \text{for } \lambda = 0.768~\micron
        \end{array}\right\}
  \]

  Table-\ref{tab1} \& \ref{tab2} shows all the details of different parameters used in Model$_X$.

\subsubsection{Model$_R$: constant monomer radius} \label{modelr}
In this modelling approach we consider fixed value of primary particle (monomer) radius or mean monomer radius \textit{r$_p$} = 0.073$\micron$ for all the three wavelengths, while the aggregate radius (\textit{R$_a$}) is considered to be in the range 0.22 to 2.0 $\micron$ for all the wavelengths (see Table-\ref{tab3} for more details). On the other hand the radius/size (\textit{R}) of solid particles are also fixed for all the wavelengths. The minimum and maximum radii of PP (monomers) and the related size parameters for the three respective wavelengths are shown below,

  \[
    r_p = \frac{\lambda}{2\pi}x_p = 0.073\micron\left\{\begin{array}{lr}
        x_p = 0.82, & \text{for } \lambda = 0.557~\micron\\
        x_p = 0.70, & \text{for } \lambda = 0.655~\micron\\
        x_p = 0.60, & \text{for } \lambda = 0.768~\micron
        \end{array}\right\}
  \]
The upper and lower size cuttoffs of monomer size under Model$_R$ is defined using lognormal size distribution having standard deviation of $\sigma_p$ = $\pm$ 0.03$\micron$. Thus the smallest monomer size is 0.043$\micron$ and 0.103$\micron$ for all the three wavelengths.
For the low porous solid particles, 50 structures are generated in the size range 0.07$\micron$ to 2.0$\micron$ for all the wavelengths. The minimum and maximum radii of Solids and the related size parameters for the three respective wavelengths are shown below,

  \[
    R_{min} = \frac{\lambda}{2\pi}X = 0.07\micron\left\{\begin{array}{lr}
        X = 0.79, & \text{for } \lambda = 0.557~\micron\\
        X = 0.67, & \text{for } \lambda = 0.655~\micron\\
        X = 0.57, & \text{for } \lambda = 0.768~\micron
        \end{array}\right\}
  \]

  \[
    R_{max} = \frac{\lambda}{2\pi}X = 2.0\micron\left\{\begin{array}{lr}
        X = 22.5, & \text{for } \lambda = 0.557~\micron\\
        X = 19.2, & \text{for } \lambda = 0.655~\micron\\
        X = 16.4, & \text{for } \lambda = 0.768~\micron
        \end{array}\right\}
  \]  
  Table-\ref{tab4} shows all the details of different parameters used in Model$_R$.

\section{Results}
In this section, we discuss the results from light scattering simulations of RFA structures for the aggregate samples (1-6) from the Granada Amsterdam Light Scattering Database. Finally, we explain the RFA+SOLID model results used to model the observed polaeization from the comet 2I/Borisov.

\subsection{Validating RFA model structures}

To replicate the morphology of cosmic dust aggregates, we generate the polydisperse RFA model structures using the package REST \citep{Halder2022a}. REST takes the structure file of a polydisperse fractal aggregate (FA) and crafts roughness and/or irregularities on the surface of each spherical grain of a FA structure and thereby creating the most realistic computer modelled cosmic dust analog, as shown in Figure-\ref{fig:1}. To proceed further, it is necessary to cross-check whether the RFA Model structures are reliable enough to be considered as pristine cosmic dust candidate to model the observed polarization of the comet 2I/Borisov. Thus, to validate the RFA Model structures light scattering simulations are performed for each of the RFA Model structures using the discrete dipole approximation (DDSCAT) to extract the light scattering parameters S11 (phase function), -S12/S11 (degree of linear polarization) and S22/S11 ratio for the respective model structures, where S(ij) are the orientationally symmetric scattering matrix elements. The simulations are performed considering the similar values of monomer size, aggregate size, refractive index and porosity as provided in the Granada Amsterdam Light Scattering Database for the aggregate samples (1-6) (see Table-\ref{tab5}). Figure-\ref{fig:3} shows the variation of -S12/S11 and S11 [normalized by S11(30$^{\circ}$) as in case of experiments] with scattering angle for RFA model structures (1-6) compared with the experimental results for Aggregate Samples (1-6) from the Granada Amsterdam Light Scattering Database shows the variation of the degree of linear polarization with the scattering angle for the different RFA model structures and those obtained from the experiments. Figure-\ref{fig:4} shows variation of S22/S11 with scattering angle for RFA model structures (1-4) compared with those which are experimentally obtained for Aggregate Samples (1-4). It is clear from the figures that when roughness is induced on the surface of spherical monomers of a fractal aggregate, a better agreement with the experimental data can be achieved. Hence, the RFA Model structure not only look similar to original cosmic dust aggregates but can also produce light scattering response similar to those obtained from light scattering experiments over cosmic dust analogs.

\begin{figure*}
    \centering
   % \vspace{10cm}
    \includegraphics[scale=0.15]{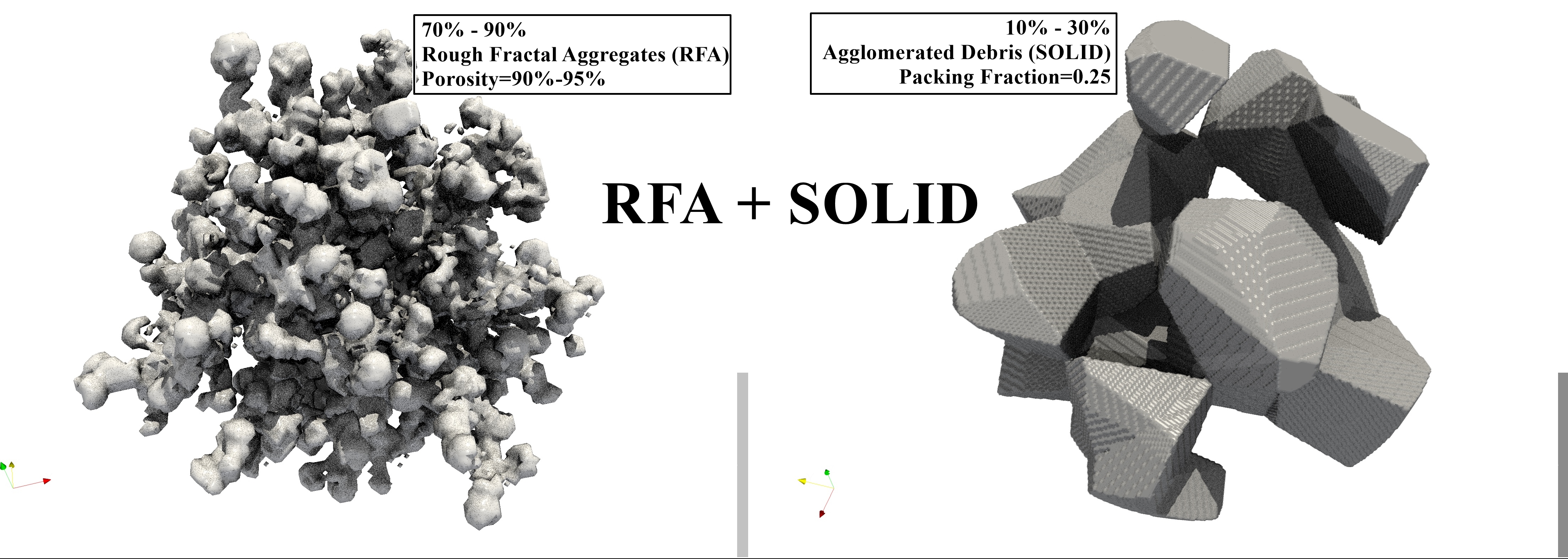}
    \caption{3D visualization of RFA + Solid mixed morphology used to model the observed polarization-phase data of comet 2I/Borisov.}
    \label{fig:5}
\end{figure*}

\begin{figure*}
    \centering
   % \vspace{10cm}
    \includegraphics[scale=0.85]{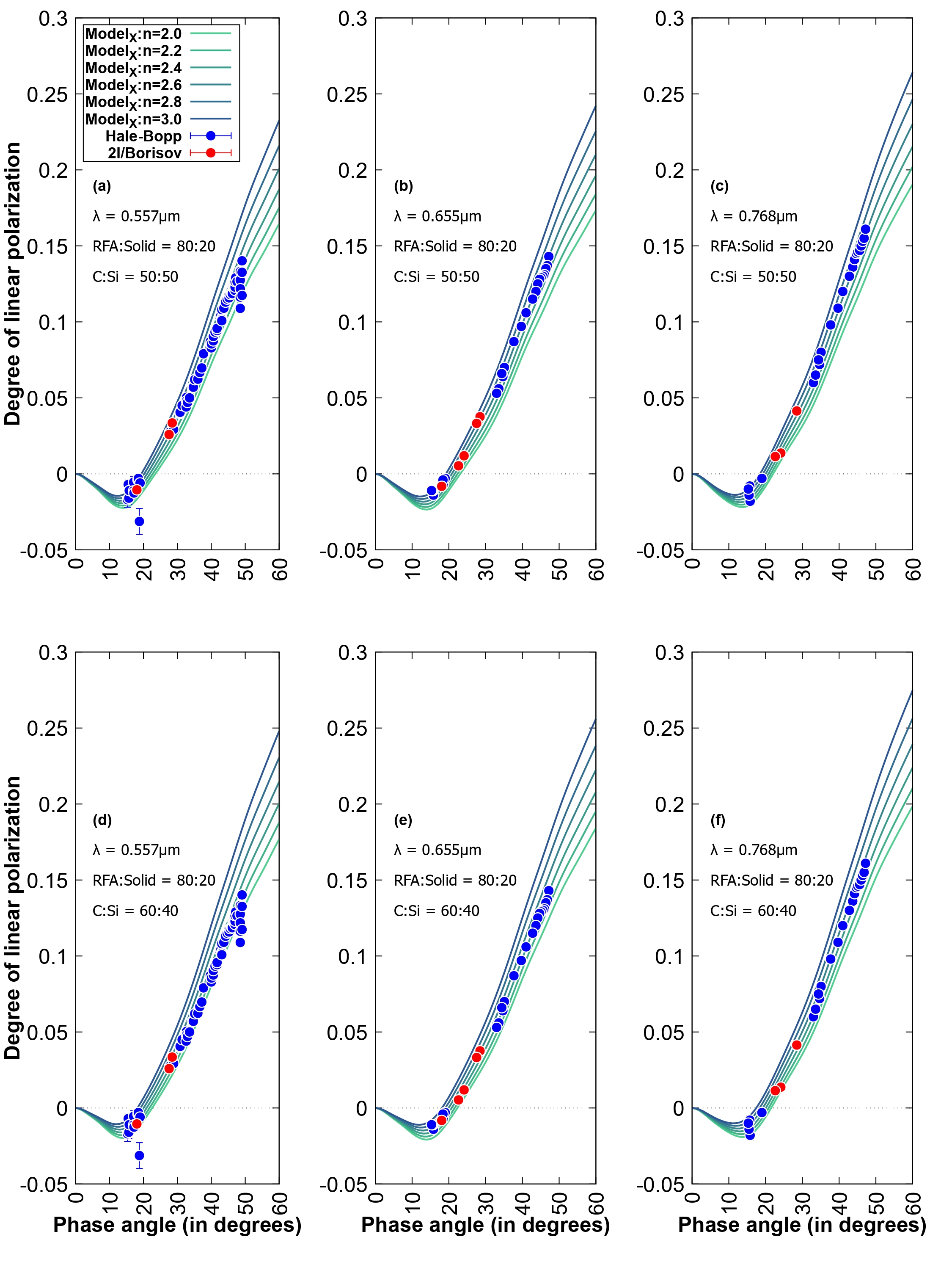}
    \caption{Variation of the degree of linear polarization with phase angle using Model$_X$ for mixed morphology RFA:Solid = 80:20 over the power-law index range \emph{n} = 2.0 - 3.0 having C:Si = 50:50 (a-c) and C:Si = 60:40 respectively for the wavelengths $\lambda$ = 0.557$\mu$m ($V_f$ filter), 0.655$\mu$m ($R_f$ filter) and 0.768$\mu$m ($I_f$ filter).}
    \label{fig:6}
\end{figure*}

\subsection{Modelling the observed polarization of 2I/Borisov}
After validating the RFA model structures, we proceed further to develop a comet dust model considering both porous cosmic dust aggregates (RFA) and low porous solid particles (Solids) (see Figure-\ref{fig:5}). In this model, the RFA structures are considered to represent the porous cosmic dust aggregates, while the agglomerated debris particles are considered to represent low porous solid particles. The recent study on modelling of cometary polarization by \cite{Halder2021b} shows a detailed modelling technique where relatively larger particles are used to model both the short and long period comets using hierarchical aggregates, fluffy solids and agglomerated debris particles. Although, the model is able to explain the observed polarization of short period comets 1P/Halley, 67P/C-G \citep{Halder2021b} and 156P/Russel-Linear \citep{Aravind2022}, but it showed certain discrepancies in the negative polarization in case of the comet Hale-Bopp. Thus, it is clear that the third class of comets require a special treatment with much simpler approach. Hence in the present study we consider relatively smaller size particles $\leq$ 2.5µm with more pristine morphology. 

We have used DDSCAT to compute the degree of linear polarization for the high porous RFA model structures and the low porous Solid particles under the parameterization schemes of Model$_X$ and Model$_R$ discussed in the Section-\ref{sec:dustmodel}. It is clear from the Rosetta/MIDAS findings that both high porous aggregates and low porous solids are present in a comet. Also, the Carbon to Silicate ratio or high absorbing material to low absorbing material ratio was found to be 50:50. Initially we mixed the simulated polarization data of porous RFA structures for amorphous silicate and amorphous carbon under the mixing ratio C:Si = 60:40 and 50:50. Finally, we mixed the inhomogeneous RFA polarization data with Solid silicate data under the mixing ratios RFA:Solid = 80:20. The best-fit results obtained under both the schemes are explained below.

\subsubsection{Best-fit results using Mode$_X$}
Figure-\ref{fig:6} depicts the variation in the degree of linear polarization using Model$_X$ for RFA:Solid = 80:20 with varying power-law index n under the different wavelength filters $\lambda$ = 0.557$\mu$m ($V_f$ filter), 0.655$\mu$m ($R_f$ filter) and 0.768$\mu$m ($I_f$ filter) respectively for C:Si=50:50 (a-c) and C:Si = 60:40 (d-f). These figures portray a multi-dimensional approach of the model where we compare the observations of 2I/Borisov and Hale-Bopp for the particular power-law index over all the three wavelengths. One can easily notice from all the three figures that the polarimetric observations of the comets 2I/Borisov and Hale-Bopp show good agreement with model curves in the power-law index range of 2.4 to 2.8 in all the three wavelengths. Although the power-law index must remain consistant over all the three wavelengths. This discrepancy may arise due to the consideration of fixed monomer size parameter which is the basis of Model$_X$. This issue is resolved when fixed monomer size is considered over all the three wavelengths as explained in the next section. 
%Considering this, we found the best-fit results having power-law index 2.7 when C:Si=50:50 and 2.5 when C:Si=60:40. 
%Figure-\ref{fig:7} (a-c) \& Figure-\ref{fig:7} (d-f) shows the best-fit curves obtained from the model for the comet 2I/Borisov at $\lambda$ = 0.557$\mu$m ($V_f$ filter), 0.655$\mu$m ($R_f$ filter) and 0.768$\mu$m ($I_f$ filter) respectively for C:Si = 50:50 and 60:40. The best fit results using Model$_X$ indicate presence of 70 - 90\% of porous RFA structures 10 - 20\% of Solid (low porosity) particles having an overall power-law size distribution index of 2.7 (C:Si=50:50) and 2.5 (C:Si=60:40). Hence, the model indicates presence of highly porous aggregates and relatively smaller size particles in the coma of comet 2I/Borisov.

\begin{figure*}
    \centering
    \vspace{-1cm}
    \includegraphics[scale=0.165]{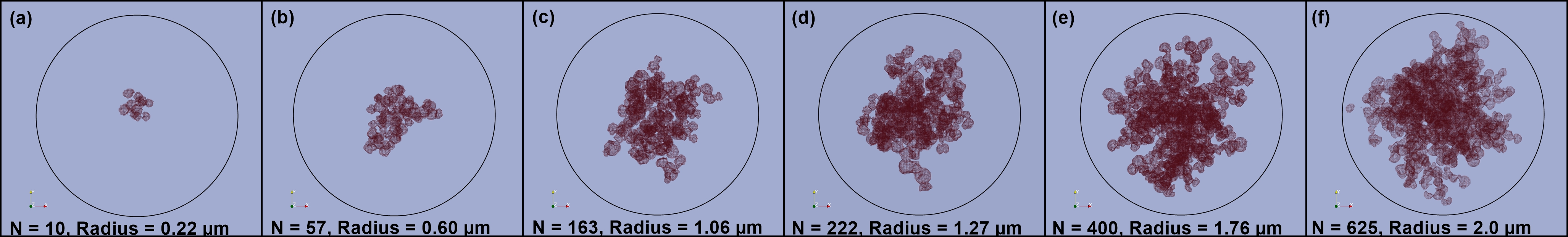}
    \caption{Visualisation of six out of 50 RFA structures for Model$_{R}$ having radii (\textit{R$_{a}$}) (a) \textit{R$_{a}$} = 0.22$\micron$ (b) \textit{R$_{a}$} = 0.60$\micron$ (c) \textit{R$_{a}$} = 1.06$\micron$ (d) \textit{R$_{a}$} = 1.27$\micron$ (e) \textit{R$_{a}$} = 1.76$\micron$ (f) \textit{R$_{a}$} = 2.0$\micron$.}
    \label{fig:7}
\end{figure*}

\begin{figure*}
    \centering
    %\vspace{-6cm}
    \includegraphics[scale=0.85]{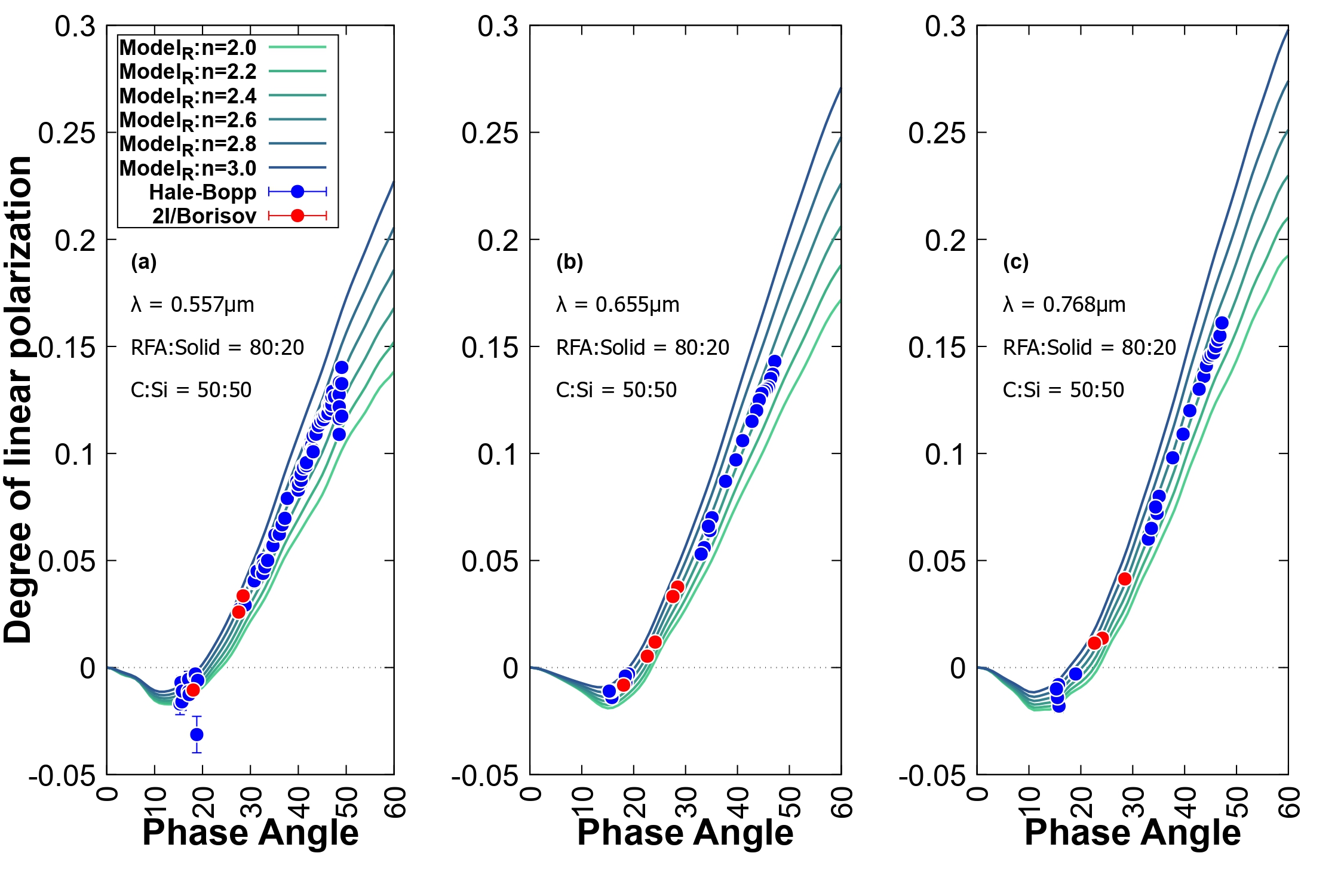}
    \caption{Variation of the degree of linear polarization with phase angle using Model$_R$ for mixed morphology RFA:Solid = 80:20 over the power-law index range \emph{n} = 2.0 - 3.0 having C:Si = 50:50 (a-c) for the wavelengths $\lambda$ = 0.557$\mu$m ($V_f$ filter), 0.655$\mu$m ($R_f$ filter) and 0.768$\mu$m ($I_f$ filter)}
    \label{fig:8}
\end{figure*}

\begin{figure*}
    \centering
   % \vspace{10cm}
    \includegraphics[scale=0.85]{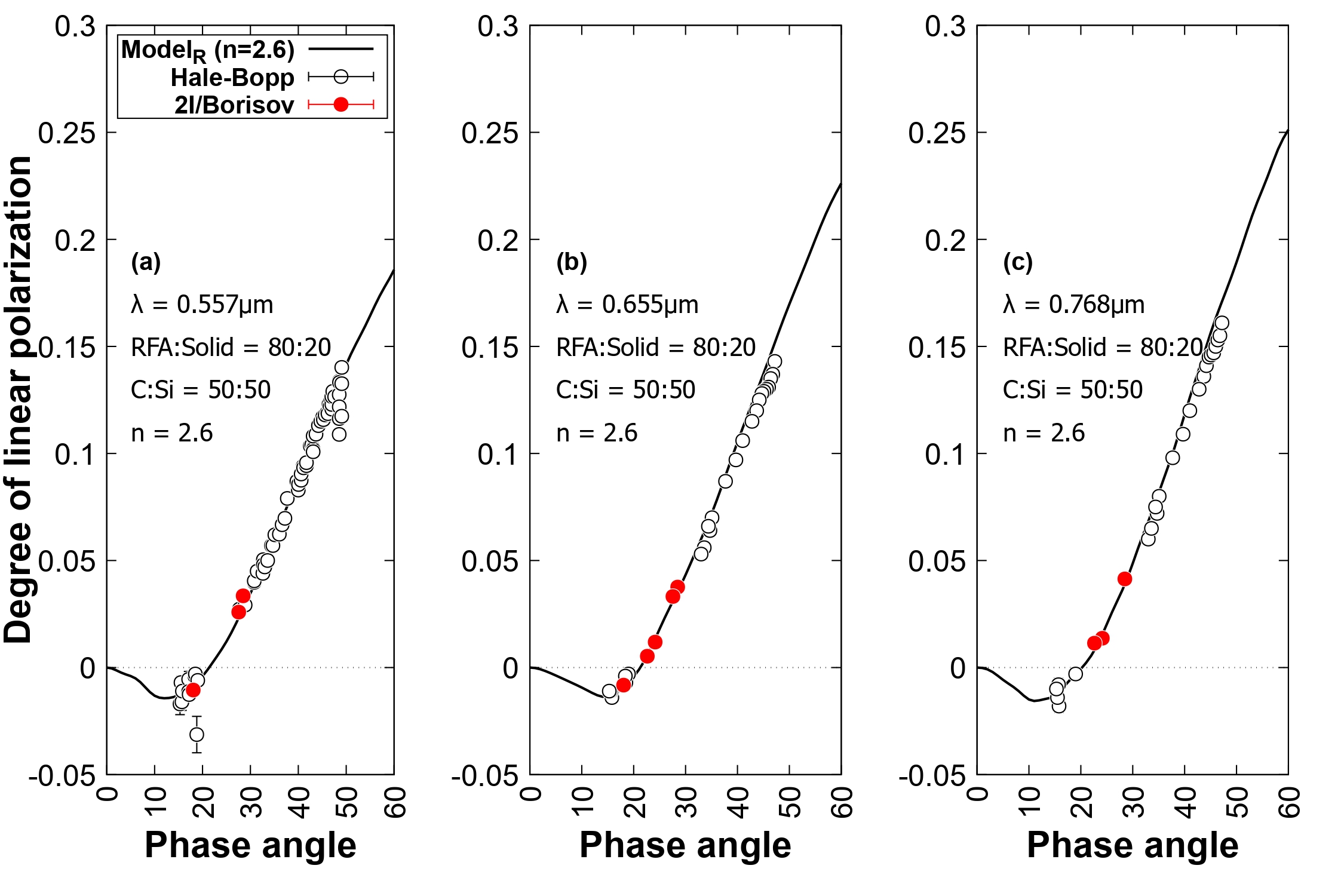}
    \caption{Best fit model results using Model$_R$ for the interstellar comet 2I/Borisov with the observations (red filled circles) at wavelengths $\lambda$ = 0.557$\mu$m ($V_f$ filter), 0.655$\mu$m ($R_f$ filter) and 0.768$\mu$m ($I_f$ filter) and compared with the polarimetric observations of the comet C/1995 O1 (Hale-Bopp) (blue filled circles) at $\lambda$ = 0.4845$\mu$m \citep{Ganesh1998PolarimetricHale-Bopp}, 0.620$\mu$m \citep{Ganesh1998PolarimetricHale-Bopp} and 0.730$\mu$m \citep{Kikuchi2006LinearComets} for C:Si = 50:50 (\emph{n}=2.6) (a-c).}
    \label{fig:9}
\end{figure*}

\subsubsection{Best-fit results using Mode$_R$}
In case of Model$_X$, the monomer size parameter is fixed for all the three wavelengths, the monomer radii is scaled according to the size parameter. But in a realistic case the monomer radii shall remain constant for the smaller and larger aggregates over all the wavelengths. This anomaly is corrected in Model$_R$ which considers fixed value of monomer radii for all the aggregates in all the three wavelengths. In this case study the variation of the degree of linear polarization and phase angle for changing aggregate sizes keeping the monomer radii fixed in all the three wavelengths for silicate and carbon respectively for six out of 50 RFA structures (see Figute-\ref{fig:7}) and 50 Solid particles following the parameterization scheme of Model$_R$ explained in Section-\ref{modelr}. Figure-\ref{fig:8} depicts the variation in the degree of linear polarization using Model$_R$ for RFA:Solid = 80:20 with varying power-law index \textit{n} under the different wavelength filters $\lambda$ = 0.557$\mu$m ($V_f$ filter), 0.655$\mu$m ($R_f$ filter) and 0.768$\mu$m ($I_f$ filter) respectivly for C:Si=50:50 (a-c). It is clear from the figure that the best-fit size distribution index $n$ remains consistant at 2.6 which becomes more clear from  Figure-\ref{fig:9} shows the best-fit curves at \textit{n}=2.6 obtained using Model$_R$ for the comet 2I/Borisov at $\lambda$ = 0.557$\mu$m ($V_f$ filter), 0.655$\mu$m ($R_f$ filter) and 0.768$\mu$m ($I_f$ filter) respectively for C:Si = 50:50.

\begin{figure*}
    \centering
   % \vspace{10cm}
    \includegraphics[scale=0.48]{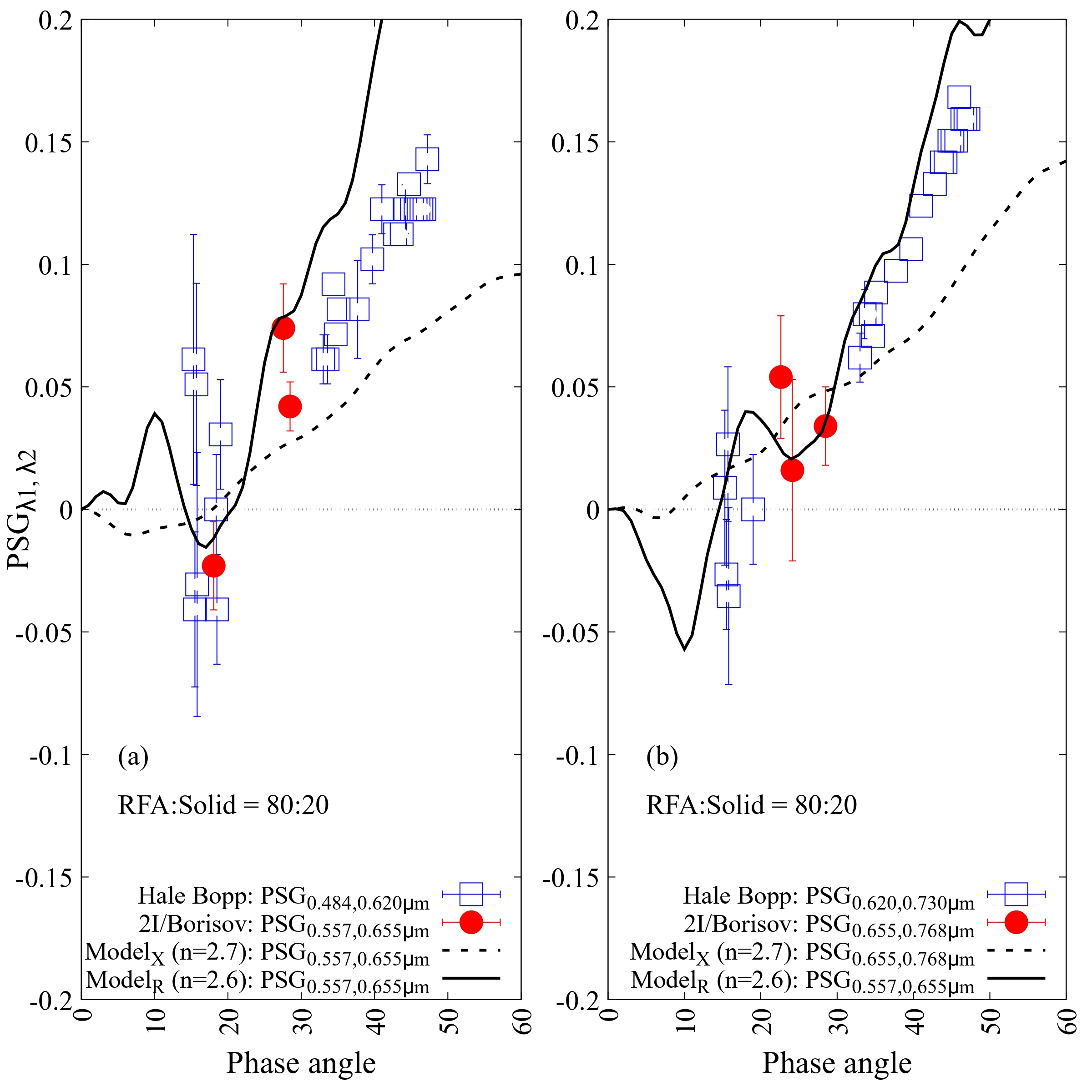}
    \caption{Best-fit results of polarimetric spectral gradient for the two pairs of wavelengths (a) PSG$_{0.557,0.655}$ and (b) PSG$_{0.655,0.768}$ (h) using Model$_X$ (dashed line) and Model$_R$ (solid line) over the observed PSG values for Hale-Bopp (blue hollow square) and 2I/Borisov (red solid circle)}
    \label{fig:10}
\end{figure*}

\begin{figure*}
    \centering
   % \vspace{10cm}
    \includegraphics[scale=0.48]{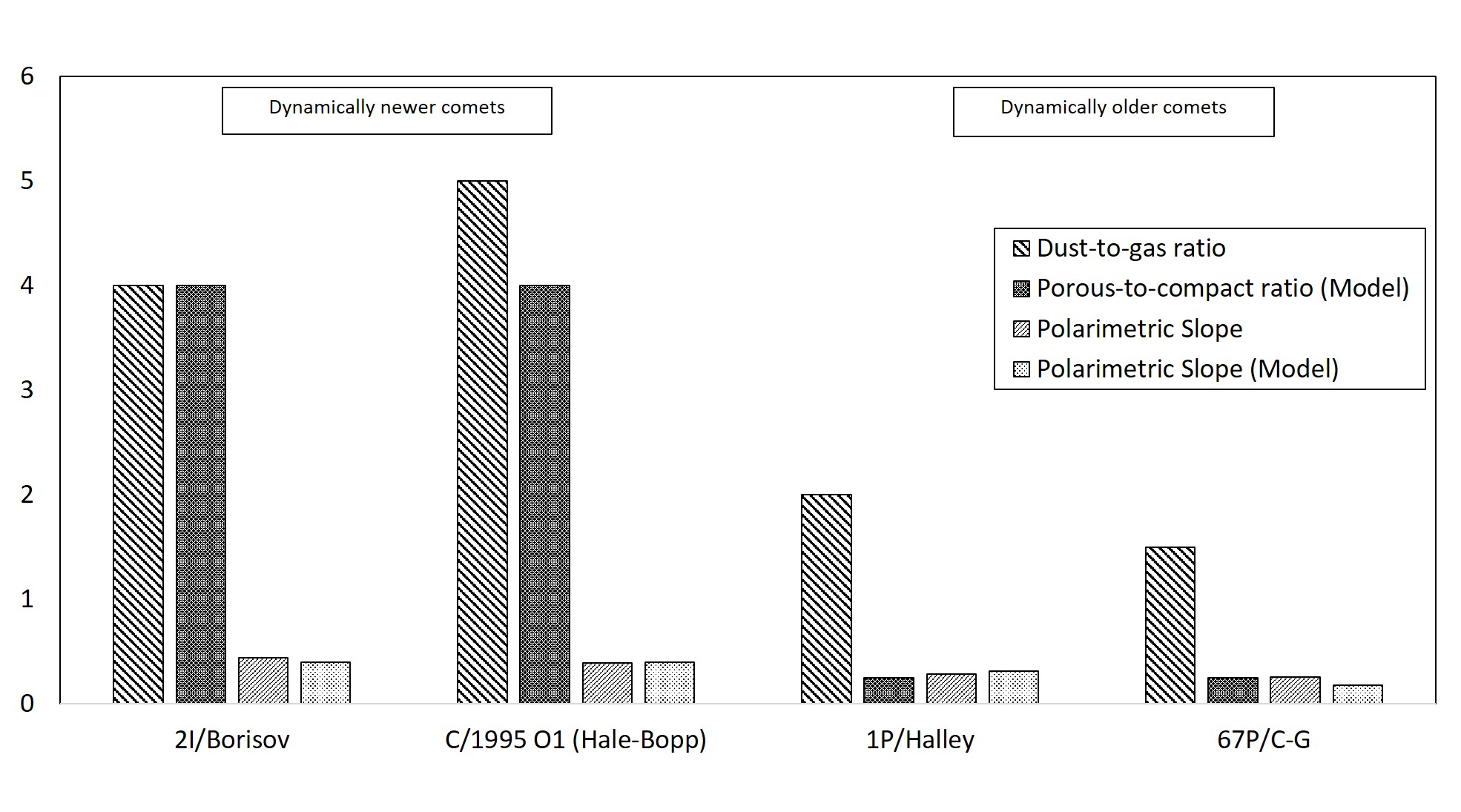}
    \caption{Comparison of \emph{dust-to-gas} ratio (measured) \citep{Jewitt1999ParticulateHale-Bopp, Yang2021Compact2I/Borisov, McDonnell1991PhysicalDust, Biver2019Long-termInstrument}, \emph{porous-to-compact} ratio (modelled) and polarimetric slope (observed \& modelled) for dynamically newer and older comets.}
    \label{fig:11}
\end{figure*}

\subsection{Modelling the Polarimetric Spectral Gradient (PSG)}
In this section we discuss the observed and modelled Polarimetric Spectral Gradient (PSG) which is defined by the following equation,

\begin{equation}
    \frac{dP(\lambda)}{d\lambda} \simeq PSG_{(\lambda_{1}, \lambda_{2})} = \frac{P(\lambda_{2}) - P(\lambda_{1})}{\lambda_{2} - \lambda_{1}}
\end{equation}
where \textit{P} is the value of polarization (observed/modelled), while $\lambda_{1}$ and $\lambda_{2}$ are the two subsequent wavelengths.
The polarimetric observations of the comet 2I/Borisov indicate that in the positive branch, polarization increases with increasing wavelength and hence the polarization spectral gradient (PSG) remains positive. The model PSG curves shown in Figure-\ref{fig:10} under Model$_X$ and Model$_R$ indicates a similar trend which is a common feature for all comets including the third class of comets having polarization higher than the high polarization comets. But in the negative branch, the observed PSG becomes negative at around 20$^\circ$ phase angles for both Hale-Bopp and 2I/Borisov. Surprisingly, the modelled PSG also becomes negative around 20$^\circ$ phase angle. Although the model PSG($V_f$,$R_f$) curve obtained using Model$_x$ indicate a significant negative trend in low phase angles, but it does not show a promising fit with the observed PSG points. Moreover, the model PSG($R_f$,$I_f$) curve obtained using Model$_X$ does not indicate any significant negative trend which is observed in case of Hale-Bopp. 
On the other hand, the model PSG($V_f$,$R_f$) and PSG($R_f$,$I_f$) obtained using Model$_R$ show significant negative trend in the low phase angles and also the model curves almost fit with some of the observed points and remains relatively close to rest of the observed values. Hence, Model$_R$ which considers fixed value of monomer radii in all the three wavelengths produces better results, while Model$_X$ which considers fixed monomer size parameter is physically incorrect and it is clear from the PSG plot. 
%The model PSG curves shown in Figure-\ref{fig:5} (g-h) indicates a similar trend which is a common feature for almost all comets including the third-class of comets having polarization higher than that of high polarization comets. But in the negative branch an unusual drop in polarization is observed in the $R_f$ filter, due to which the PSG($V_f$, $R_f$) becomes negative below 20$^{\circ}$. Surprisingly, the modelled PSG($V_f$, $R_f$) also becomes negative below 20$^{\circ}$ phase angle. On the other hand the modelled PSG($R_f$, $I_f$) show negligibly small negative value below 10$^{\circ}$ phase angle. In the positive branch both the modelled PSG remains positive as those obtained from the observations. 

\subsection{Exploring relation between \emph{dust-to-gas} ratio and intrinsic dust parameters}
The best-fit model data for the observed polarization and PSG of the third class of comets 2I/Borisov \& Hale-Bopp and those of short period comets 1P/Halley and 67P/C-G obtained from the light scattering simulations explained in this study and \cite{Halder2021b} respectively, indicate that the intrinsic properties of dust play a crucial role in defining the signature polarimetric slope of different class of comets. For example, in this study the \textit{porous-to-compact} ratio (RFA:Solid) obtained for the comets 2I/Borisov \& Hale-Bopp (third-class of comets) is 4. While the \textit{porous-to-compact} ratio (HA:Solid) obtained for comets 1P/Halley \& 67P/C-G (short period comets) is 0.25. Thus, the difference in polarization or the polarimetric slope of different class of comets tends to be proportional to the \textit{porous-to-compact} ratio of dust in the coma of the comet. In a similar way, the \textit{dust-to-gas} ratio in the coma of a comet is proportional to the polarization or polarimetric slope. Figure-\ref{fig:11} depicts the respective \textit{dust-to-gas} ratio and \textit{porous-to-compact} ratio for the aforesaid comets. It is clear from the figure that high \textit{dust-to-gas} ratio in the third class of comets is accompanied by high \textit{porous-to-compact} ratio of dust particles, while, in low polarization comets the \textit{dust-to-gas} ratio is accompanied by low \textit{porous-to-compact} ratio.  

\section{Discussion}
Under the framework of polarimetric observations of the interstellar comet 2I/Borisov \citep{Bagnulo2021} and the light scattering experiments of different aggregate samples of cosmic dust analogs from the Granada Amsterdam Light Scattering Database \citep{Volten2007AstrophysicsAnalogs} we develop a visually realistic dust model to replicate the unusual polarization phase curve observed in the comet 2I/Borisov. We obtained the best fit model for a mixture of porous RFA particles 80\% and Solids 20\% having power-law size distribution index $n$ = 2.7 using parameterization scheme of Model$_X$ and $n$ = 2.6 using parameterization scheme of Model$_R$ respectively for C:Si = 50:50 under wavelength filters $\lambda$ = 0.557$\mu$m ($V_f$ filter), 0.655$\mu$m ($R_f$ filter) and 0.768$\mu$m ($I_f$ filter). The higher percentage of porous RFA structures and the higher power-law index indicate that the coma of 2I/Borisov is dominated by porous and relatively smaller dust particles. On the other hand, best fit model results for short period comets 1P/Halley, 67P/Churyumov-Gerasimenko \& 156P/Russel-LINEAR \citep{Halder2021b, Aravind2022} indicate presence of lesser amount of porous aggregates. Also, the study indicates that the \textit{porous-to-compact} ratio of dust particles is directly proportional to the \textit{dust-to-gas} ratio observed in the coma of a comet. In case of relatively newer comets, the coma is dominated by high amount of small size porous dust aggregates of Rayleigh size grains indicating high \textit{dust-to-gas} ratio owing to high \textit{porous-to-compact} ratio thereby producing relatively higher polarization. On the other hand in case of short-period or older comets, the coma is dominated by gas and large size compact dust particles that are mainly concentrated in the inner coma and the near nucleus regions indicating low \textit{dust-to-gas} ratio owing to low \textit{porous-to-compact} ratio which in turn produces lower polarization. Thus, it is very much clear from this study that dynamically new comets carry larger amount of porous pristine cosmic dust particles, while in dynamically older comets or short period comets a larger portion of the pristine dust particles are lost due to frequent weathering by Solar wind. In this study we employ a multi-dimensional approach by considering morphologically realistic dust particles having a mixture of high and low porosity over a wide range of size distribution for the three broadband filters. We all know 2I/Borisov as an interstellar comet, yet it was once part of an extrasolar planetary system and hence it was an exocomet before drifting away from its host star. Astronomers have recently found signatures of exocomets using techniques such as photometric transits and far IR/mm gas emission from within debris disks \citep{Strm2020ExocometsPerspective, LecavelierdesEtangs2022ExocometsSystem}. On the other hand recent observations of main-belt comets (active asteroids) reveal certain amount of dust and gas production rates \citep{Jewitt2012THEASTEROIDS, Moreno2021DustPANSTARRS}. Hence, the correlation between \textit{dust-to-gas} ratio and \textit{porous-to-compact} ratio indicated in this study can be of great use to determine the intrinsic dust properties in main-belt comets and exocomets. The model has been verified considering best-fit results for the observed polarization and polarimetric spectral gradient of the interstellar comet 2I/Borisov. This study ensures that the RFA model structures represent the pristine morphology of cosmic dust particles and are capable enough to reproduce experimental as well as observational data. Hence, these structures will be highly useful for future studies related to cometary dust polarization, polarimetric response from protoplanetary disks, atmosphere of cloudy exoplanets and brown dwarfs \citep{Chakrabarty2022PolarizationAtmospheres, Marley2011ProbingPolarization, Sengupta2001PROBINGPOLARIZATION}, extinction of background starlight in dense molecular clouds and polarimetric study of dust in the circumstellar environments. Although, we tried to develop a realistic cosmic dust model considering realistic dust particles having surface roughness/irregularities, yet the model can be improved by considering large size hierarchical aggregates and including compositions such as organics, FeS and different kinds of ices. These limitations can be addressed in some future work to develop a more realistic and generalised comet dust model. 

\section{Acknowledgements}
The authors deeply thank the anonymous reviewers for their fruitful suggestions, which has enriched the manuscript with greater details. The authors acknowledge the high-performance computing facility (NOVA) of the Indian Institute of Astrophysics, Bangalore, where all the intensive light scattering simulations are conducted. The authors also acknowledge Prof. Shashikiran Ganesh of PRL, Ahmedabad and Dr. Himadri Sekhar Das of Dept. of Physics, Assam University, Silchar for important discussions.

\newpage

\appendix

\section{RFA Model Structures}
\begin{figure*}
    \centering
   % \vspace{10cm}
    \includegraphics[scale=0.15]{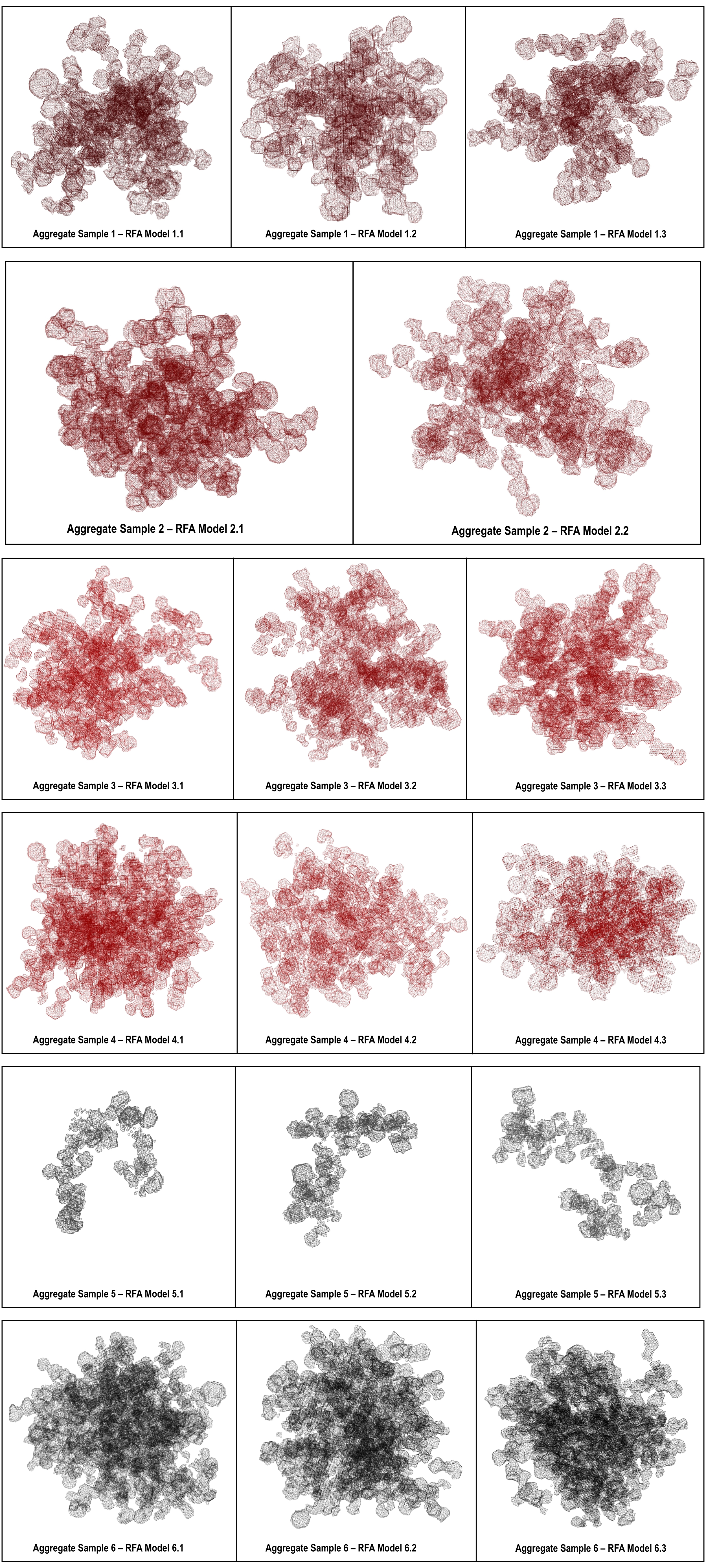}
    \caption{RFA Model structures (1-6) representing Aggregate Samples (1-6) from the Granada Amsterdam Light Scattering Database.}
    \label{fig:A1}
\end{figure*}

The RFA model structures used in this study are the exact computer modelled replica of circumstellar dust analogs prepared in the Amsterdam Granada Light Scattering setup. Figure-\ref{fig:A1} depicts the RFA model realisations of circumstellar dust analog samples (1-6) having similar refractive index (\textit{n}), porosity, grain size and aggregate size following Table-3 of \cite{Volten2007AstrophysicsAnalogs}. The different physical parameters of the RFA model structures are shown in Table-\ref{tab5}. For simplicity we considered the imaginary part of the refractive index $k$ = 0.0001 for all the structures, as it was not detected in the experiments. 


\bibliography{references}{}
\bibliographystyle{aasjournal}

\end{document}